\documentclass[a4paper,11pt]{article}
\pdfoutput=1 

\usepackage{jinstpub} 

\usepackage[english]{babel}
\usepackage[TS1,T1]{fontenc}
\usepackage[utf8]{inputenc}
\usepackage{siunitx}
\usepackage{todonotes}
\usepackage{mhchem}
\usepackage{booktabs}
\usepackage[export]{adjustbox}
\usepackage{subfig}
\usepackage{lineno}

\newcommand{\Cd}{\ce{^{109}Cd}}
\newcommand{\Sr}{\ce{^{90}Sr}}

\newcommand{\hmidrule}{\midrule[\heavyrulewidth]}

\begin{document}

\title{\boldmath Time resolution and power consumption of a monolithic silicon pixel prototype in SiGe BiCMOS technology}

\author[a,1]{L. Paolozzi,\note{Corresponding author.}}
\author[b]{R. Cardarelli,}
\author[a]{S. D\'ebieux,}
\author[a]{Y. Favre,}
\author[a]{D. Ferr\`ere,}
\author[a]{S. Gonzalez-Sevilla,}
\author[a]{G. Iacobucci,}
\author[c]{M. Kaynak,}
\author[d,e]{F. Martinelli,}
\author[d]{M. Nessi,}
\author[c]{H. R\"ucker,}
\author[f]{I. Sanna,}
\author[a]{DMS Sultan,}
\author[a]{P. Valerio}
\author[a,2]{and E. Zaffaroni\note{Now at EPFL.}}


\affiliation[a]{University of Geneva,\\ Geneva, Switzerland}
\affiliation[b]{INFN Section of Roma Tor Vergata,\\ Roma, Italy}
\affiliation[c]{IHP - Leibniz-Institut f\"ur innovative Mikroelektronik \\ Frankfurt (Oder), Germany}
\affiliation[d]{CERN,\\ Geneva, Switzerland}
\affiliation[e]{EPFL,\\ Lausanne, Switzerland}
\affiliation[f]{University of Torino \\ Visiting student at University of Geneva, \\ Geneva, Switzerland}

\emailAdd{lorenzo.paolozzi@unige.ch}

\abstract{ SiGe BiCMOS technology can be used to produce ultra-fast, low-power silicon pixel sensors that provide state-of-the-art time resolution even without  internal gain. The development of such sensors requires the identification and control of  the main factors that may degrade the timing performance as well as the characterisation of the dependance of the sensor time resolution on the amplifier power consumption. Measurements with a \Sr{} source of a prototype sensor produced in SG13G2 technology from IHP Microelectronics shows a time resolution of \SI{140}{\pico\second} at an amplifier current of \SI{7}{\micro\ampere} and   \SI{45}{\pico\second}  at a power consumption of \SI{150}{\micro\ampere}. The resolution on the measurement of the signal time-over-threshold, which is used to correct for time walk, is the main factor affecting the timing performance of this prototype.}





\maketitle
\flushbottom

\section{Low-power performance of SiGe HBT-based amplifiers}
\label{sec:intro}
\subsection{Time jitter comparison at low power operation of HBT and CMOS amplifiers}

The integration of pixelated  sensors in a microelectronic process has opened to the possibility of producing  sensors  with small pixels for particle-physics experiments ~\cite{alpide}~\cite{mupix} without incurring in a complex hybridization process.
An evolution of this approach is to combine a \SI{100}{\pico\second} or better time resolution to develop low-cost 4D sensors that can be produced in large volumes and assembled more rapidly and with simplified procedures.
One of the main challenges for the development of such a detector is to reduce the power consumption of the front-end electronics while maintaining excellent timing performance, in a trade-off where the limits are set by the characteristics of the transistor.

The  selection of the technology for signal amplification and read-out is also primordial.
The most direct approach to achieve excellent timing resolution with low power consumption is to use small-size transistors from down-scaled CMOS technologies.
However the cost of an engineering run in these technologies is high and the improvement in transistor speed is limited by the vulnerability of the circuit 
to its parasitic capacitances.
One alternative approach \cite{hexa_50ps} to maximize the timing performance without scaling down the process node is to use SiGe BiCMOS technologies, which offer a standard silicon CMOS process with the addition of a high-performance SiGe Hetero-junction Bipolar Transistor (HBT).
SiGe HBTs feature a transconductance and a transition frequency that can exceed those of most advanced CMOS nodes \cite{SiGe_tech}.
Moreover, they exhibit low noise for fast signal integration typical of BJT technologies \cite{GattiManfredi} and  robustness to  parasitic capacitance \cite{SiGe_tech} that make them suitable for a pixel detector.

\subsection{The SiGe BiCMOS monolithic pixel prototype}

\begin{figure}[htbp]
	\centering
	\includegraphics[width=.8\textwidth,trim=0 0 0 0]{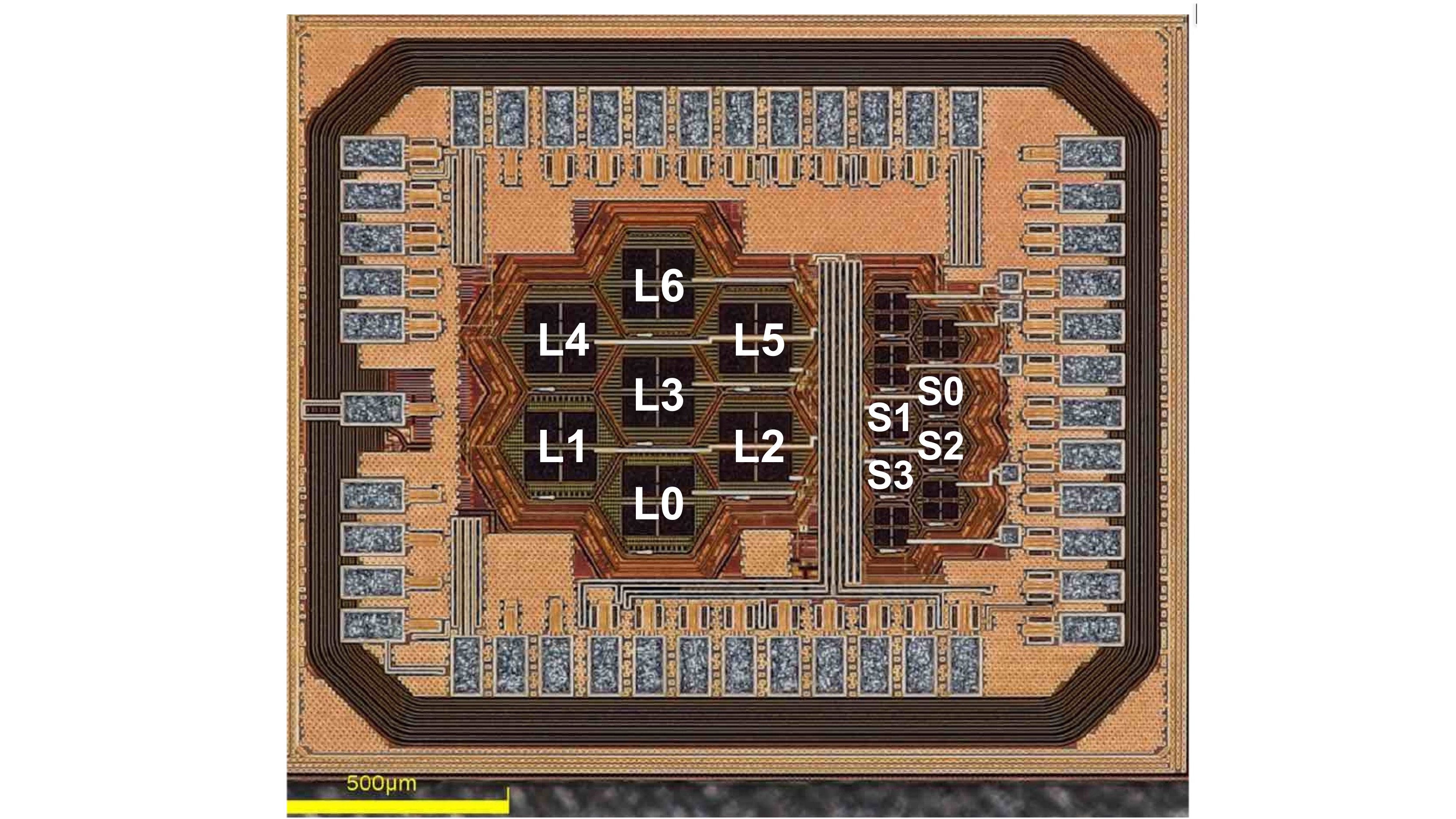}
	\caption{The monolithic prototype in IHP SG13G2 technology used for the measurements. The pixels and front-end electronics are inside deep-nwells operated at positive low-voltage, while the p-doped substrate is operated at negative high-voltage. For the measurements presented here, the output of the discriminators was sent to a differential driver and directly readout by an oscilloscope.}
	\label{fig:Hexachip}
\end{figure}

To evaluate the analogue behavior of a pixel detector with 4D tracking capability, we designed the small-size monolithic silicon pixel prototype~\cite{hexa_50ps} of Figure~\ref{fig:Hexachip} in the  IHP SG13G2 process~\cite{SG13G2} . The sensor was produced on a standard substrate resistivity of \SI{50}{\ohm\centi\meter}. It has matrices of large and small hexagonal pixels, with hexagon sides of \SI{130}{\micro\meter} and \SI{65}{\micro\meter}. Each pixel is connected to the front-end circuit, which is placed in an independent deep-nwell to insulate it from the high-voltage substrate. A study of the timing performance at high power consumption of the prototype is described in \cite{hexa_50ps}. 
In this paper we focus on the  contribution to the time resolution of the amplifier power consumption and of the time-walk correction method.

The prototype integrates a front-end electronic circuit consisting of a HBT-based amplifier and a CMOS-based discriminator originally developed to read  large pixels of a silicon sensor for medical applications \cite{TTPET_demonstrator}, which was not designed to operate at very low current. For this reason in this study the CMOS-based discriminator was kept at the original design current of \SI{40}{\micro\ampere}. The HBT-based amplifier, on the other hand, has programmable bias components and it was operated with a supply current in the range between \SI{7}{\micro\ampere} and \SI{150}{\micro\ampere}. 
The output of the discriminator was sent out from the chip using a differential CML driver, allowing the measurement of the signal Time of Arrival (ToA) and Time over Threshold (ToT). In this prototype, the measurement of the amplitude of the signal at the output of the amplifier was not possible. 
\section{Experimental setup with radioactive sources}
\label{sec:setup}
\begin{figure}[htbp]
	\centering
	\includegraphics[width=0.6\textwidth]{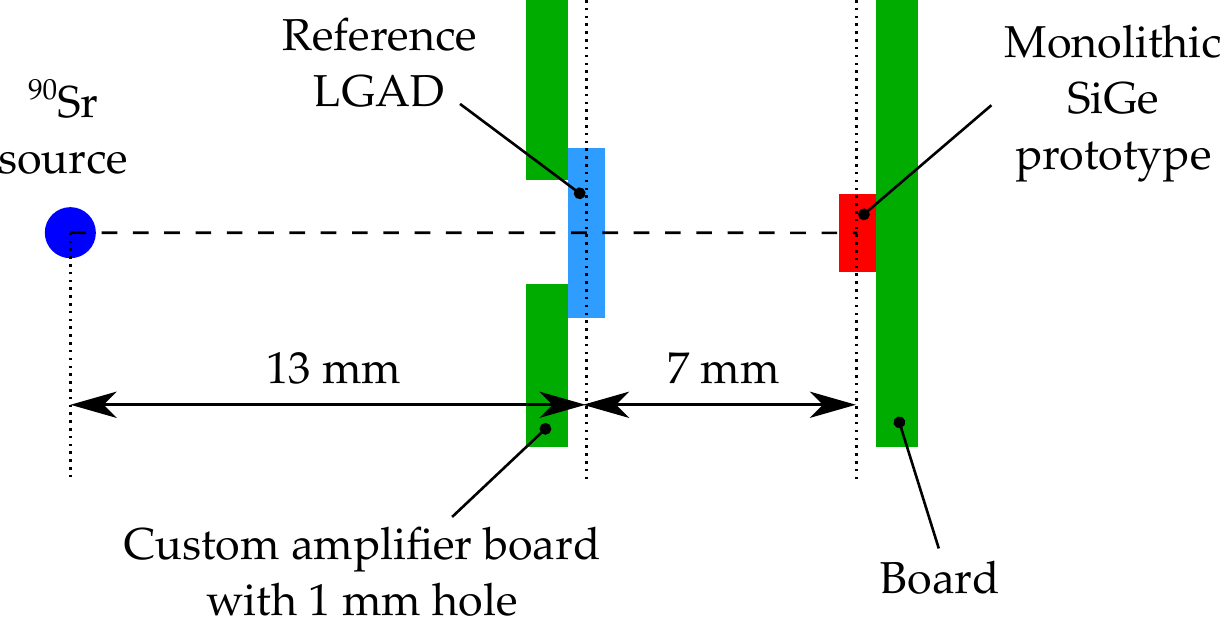}
	\caption{Sketch of the experimental setup used for the ToF measurements showing the \Sr{} source, the reference LGAD providing reference time and the monolithic prototype chip under test.}
	\label{fig:exp_setup}
\end{figure}

An experimental setup with radioactive sources (Figure~\ref{fig:exp_setup}) was built to measure the gain, noise and time resolution of the prototype chip while operating the amplifier at a collector current of \num{7}, \num{20} and \SI{50}{\micro\ampere} and the sensor at a bias voltage between \num{100} and \SI{180}{\volt}. The expected depletion depth at a bias voltage of \SI{140}{\volt} for a bulk resistivity of \SI{50}{\ohm\centi\meter} is \SI{26}{\micro\meter}, making the intrinsic sensor contribution to time resolution from the charge deposition profile and the sensor uniformity of response below \SI{30}{\pico\second}~\cite{werner}. The measurement of the gain of the amplifier at the different working points was performed using a \Cd{} source, which generates \num{22} and \SI{25}{keV} photons.

The time resolution was measured as the difference between the signal time of arrival for the pixel under test and a reference Low Gain Avalanche Diode (LGAD) using a \Sr{} source. The LGAD was glued on a dedicated amplifier board, which had a \SI{1}{mm}-wide hole to allow for the passage of the \Sr{} electrons and enable Time-of-Flight (ToF) measurement (see Figure~\ref{fig:exp_setup}). The reference LGAD used for this test had a time resolution of 50 ps RMS \cite{hexa_50ps, LGAD_FBK}. 
The board with the monolithic SiGe prototype under test was precisely mounted so to have the hole in the LGAD board aligned with the small pixel called "S0" in Figure~\ref{fig:Hexachip}.

\section{Results of the measurements}
\label{sec:results}

\subsection{Amplifier noise and gain}
\begin{figure}[h]
	\centering
	\includegraphics[width=0.7\textwidth]{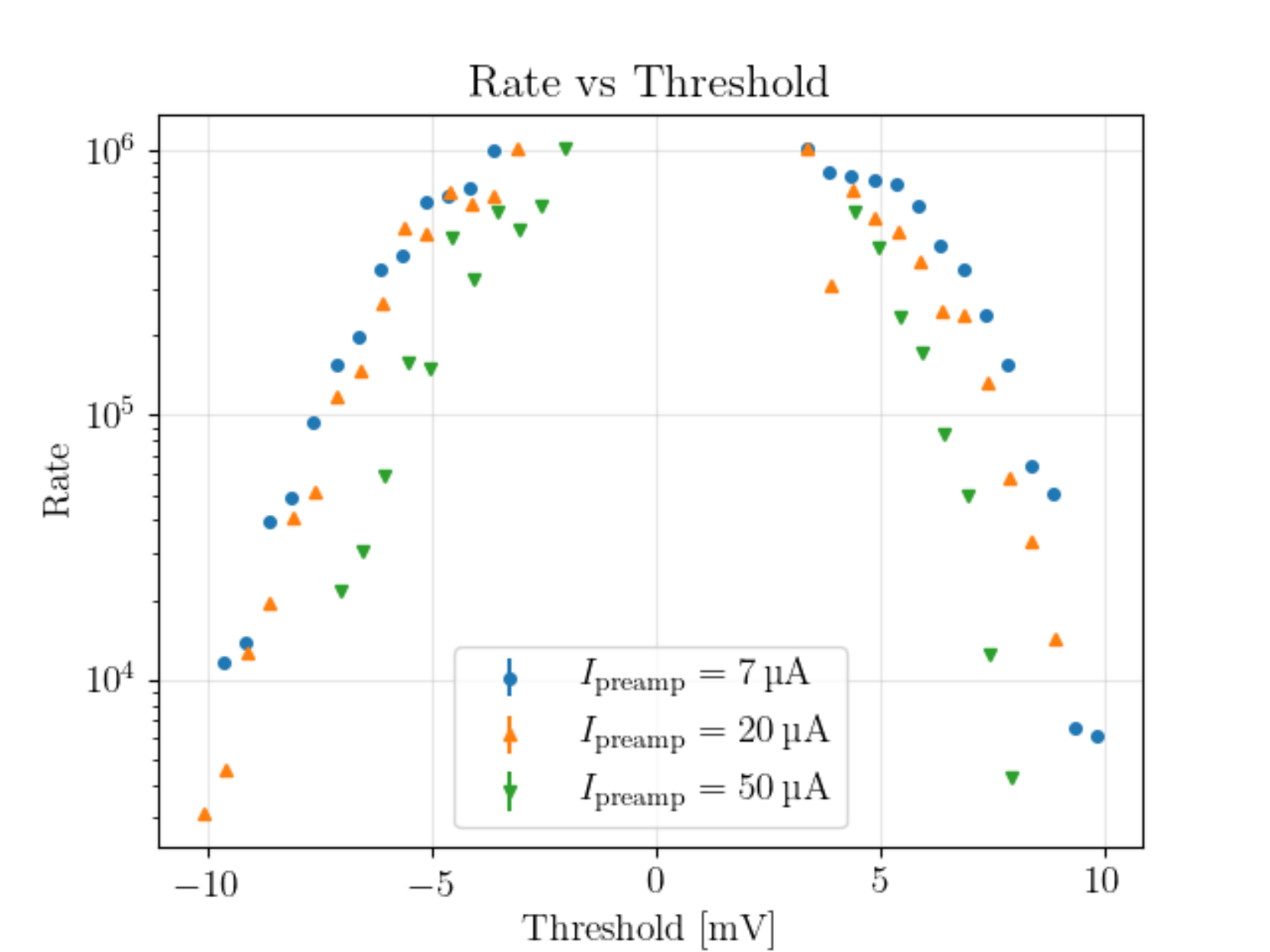}
	\caption{Noise hit rate of the pixel under test as a function of the discriminator threshold. The acquisition setup saturated at a counting rate of \SI{1}{\mega\hertz}.}
	\label{fig:noise_rate}
\end{figure}

The measurement of the noise hit rate as a function of the discriminator threshold is reported in Figure~\ref{fig:noise_rate}, for thresholds above and below the baseline and for different amplifier currents. Although the noise hit rates show a compression for values larger than \SI{500}{\kilo\hertz} due to a saturation of the counting-rate capability of the data-acquisition system, the measurements below 100kHz allow obtaining an indication of
the RMS voltage noise as seen at the output of the discriminator.

The CMOS-based discriminator acts as a filter for the amplifier noise. The ratio between the noise at the output of the amplifier ($ \sigma_V $) and the one measured after the discriminator was estimated using a Cadence Spectre\footnote{\url{https://www.cadence.com/en_US/home.html}} simulation. The simulation shows that, in order to estimate the voltage noise at the output of the amplifier that we need to calculate the Equivalent Noise Charge (ENC), the standard deviation of the voltage noise obtained by the measurement of the noise hit rate at the output of the discriminator should be increased by 20\% for an amplifier current of \SI{7}{\micro\ampere}, 50\% for a current of \SI{20}{\micro\ampere} and 60\% for a current of \SI{50}{\micro\ampere}. The resulting values after the scaling factor obtained from the simulation are reported in Table~\ref{tab:noise_gain}.

\begin{figure}[h]
	\centering
	\includegraphics[width=0.65\textwidth]{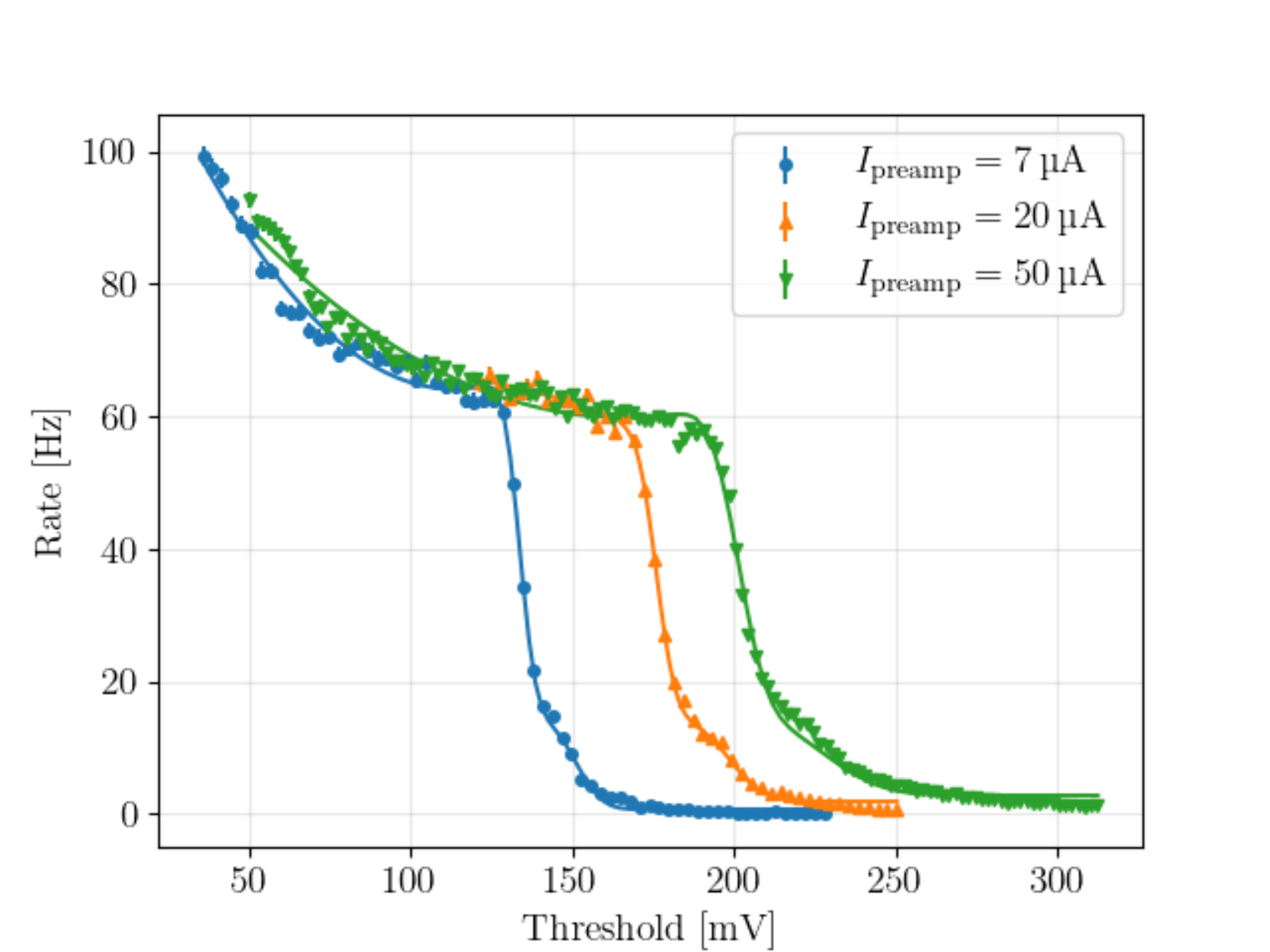}
	\caption{Photon count rate as a function of the discriminator threshold for the three values of the amplifier collector current with the sensor exposed to a \Cd{} source. The lines represent the fit done with a composition of two error functions, a constant and a second degree polynomial function; the latter was introduced to describe the rise in the low-threshold region produced by low-energy photons.}
	\label{fig:gain}
\end{figure}

The charge gain $ A_Q $ of the amplifier was estimated by measuring the photon count rate as a function of the discriminator threshold using a \Cd{} source. The results for the three amplifier currents considered are reported in Figure~\ref{fig:gain}; the data were fitted with the sum of two error functions, to account for the two main photon energies, and a constant, to account for the low-probability emission of \SI{88}{\kilo\electronvolt}  photons. A second degree polynomial was added to describe the low threshold region. The mean value of the error functions estimated by the fit was used to extract the amplifier gain values reported in Table~\ref{tab:noise_gain}. Then, the noise and gain measurements were used to estimate the Equivalent Noise Charge (ENC) values, also reported in Table~\ref{tab:noise_gain}.

\begin{table}[h]
	\centering
	\caption{Voltage noise at the output of the amplifier, charge gain and ENC for the different amplifier currents. The results for an amplifier current of \SI{150}{\micro\ampere} are those from~\cite{hexa_50ps}. The uncertainties are statistical only.}
	\label{tab:noise_gain}
	\begin{tabular}{lcccc}
		\toprule 
		$I_\text{amp}$	& Power & $\sigma_\text{V}$ & $A_Q$& ENC $= \sigma_\text{V}/A_Q$ \\
		 $[\si{\micro A}]$ & [\si{\micro\watt}] & [\si{mV}]	& [\si{mV/fC}]	& [e] \\
		\hmidrule 
		7 & $ 12.6 $ & $3.5 \pm 0.1$ & $136.6\pm0.3$ & $160\pm5$ \\ 
		\midrule
		20 & $ 36 $ & $ 4.5\pm  0.2$ & $179.5\pm0.3$ & $157\pm7$ \\ 
		\midrule
		50 & $ 90 $ & $4.2 \pm 0.3$ & $205.6\pm0.4$ & $128\pm9$ \\ 
		\midrule
		150~\cite{hexa_50ps} & $ 375 $ & $4.0 \pm 0.3$ & $290\pm2$ & $90\pm7$ \\ 
		\midrule
	\end{tabular}
\end{table}

\subsection{Time resolution}
Electrons emitted by the \Sr{} source were used to measure the jitter on the ToF between pixel "S0" and the reference LGAD detector.
The output of the discriminator of the pixel under test was read by an oscilloscope with a sampling rate of \SI{40}{GSa/s} and an analogue bandwidth of \SI{3}{GHz}.
The output signals of the LGAD were amplified using a discrete components amplifier based on SiGe HBTs~\cite{100ps_pixel} and sent to a channel of the oscilloscope.

Pixels "S1" and "S2" were operated at the lowest possible threshold clean from noise. They were attenuated and read by one channel of the oscilloscope to generate a tag signal for events with charge shared between the pixel "S0" under test and the neighbouring ones. Pixel "S3" was not used and its discriminator threshold was set to the highest possible value to prevent it from firing.

The oscilloscope trigger was configured to require a pulse from the pixel under test and a pulse from the reference LGAD within a \SI{50}{\nano\second} time window.
At each trigger, the waveforms from all channels were recorded, allowing the calculation of the ToA of all pulses, the ToT of the pixel "S0" under test and the amplitude of the LGAD signal.

During  data  analysis, each waveform was split in a signal region and a background region.
The signal region was defined as the time interval in a window of $ \pm $\SI{15}{\nano\second} around the trigger time, containing all the pulses from the LGAD and the Device Under Test (DUT).
The part of the waveform before and after the signal region --the background region-- was used to estimate the baseline and  remove noisy events: if one of the waveforms exceeded a given threshold in the background region, the event was considered to be noisy and removed from the analysis. The fraction on noisy events was  within \num{0.1}\% of the total number of events.

Coincidences between the reference LGAD and the DUT were acquired at several thresholds and bias voltages for different amplifier currents.
To keep the data analysis as simple as possible, for each measurement a unique time-walk  correction function was applied to the entire ToT range, although it was noticed that a slightly improved time resolution could be obtained by a special treatment of events at the edges of the TOT distribution. 
For the same reason, the only event selection applied to the prototype data was the removal of the very small fraction of noisy events mentioned above. 
To keep under control the reference time, only the events with LGAD pulse amplitudes in the interval in which the LGAD time resolution was  measured to be \SI{50}{ps}~\cite{hexa_50ps} were considered.

\begin{figure}[h]
	\centering
	\includegraphics[width=0.65\textwidth]{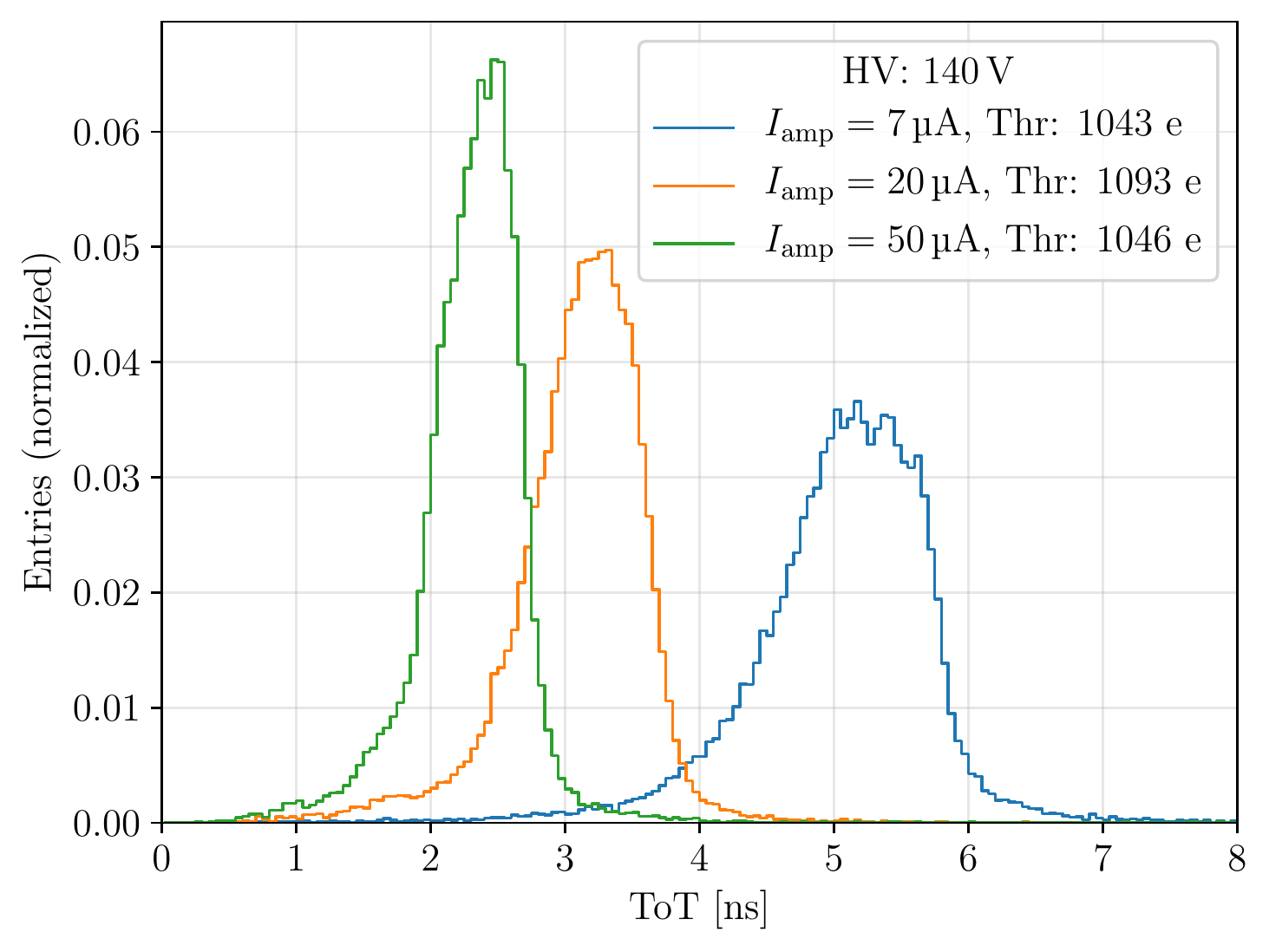}
	\caption{ToT distribution of the pixel under test for an amplifier current of \num{7}, \num{20} and \SI{50}{\micro\ampere} at a threshold of \num{1043}, \num{1093} and \num{1046} electrons, respectively, taken at a sensor bias voltage of \SI{140}{\volt}. 
	The charge-thresholds values were obtained dividing the discriminator threshold voltage by the amplifier gain.}
	\label{fig:tot_hist}
\end{figure}

Figure~\ref{fig:tot_hist} shows the distribution of the ToT for the three amplifier-current working points after the simple data analysis described. The increase of the average value of the ToT at lower currents is an indication of the reduction of the amplifier bandwidth that is in part attributed to the increase of the output impedance and in part to a reduction of the gain-bandwidth product of the transistor. 

\begin{figure}[h]
	\centering
	\includegraphics[width=0.65\textwidth]{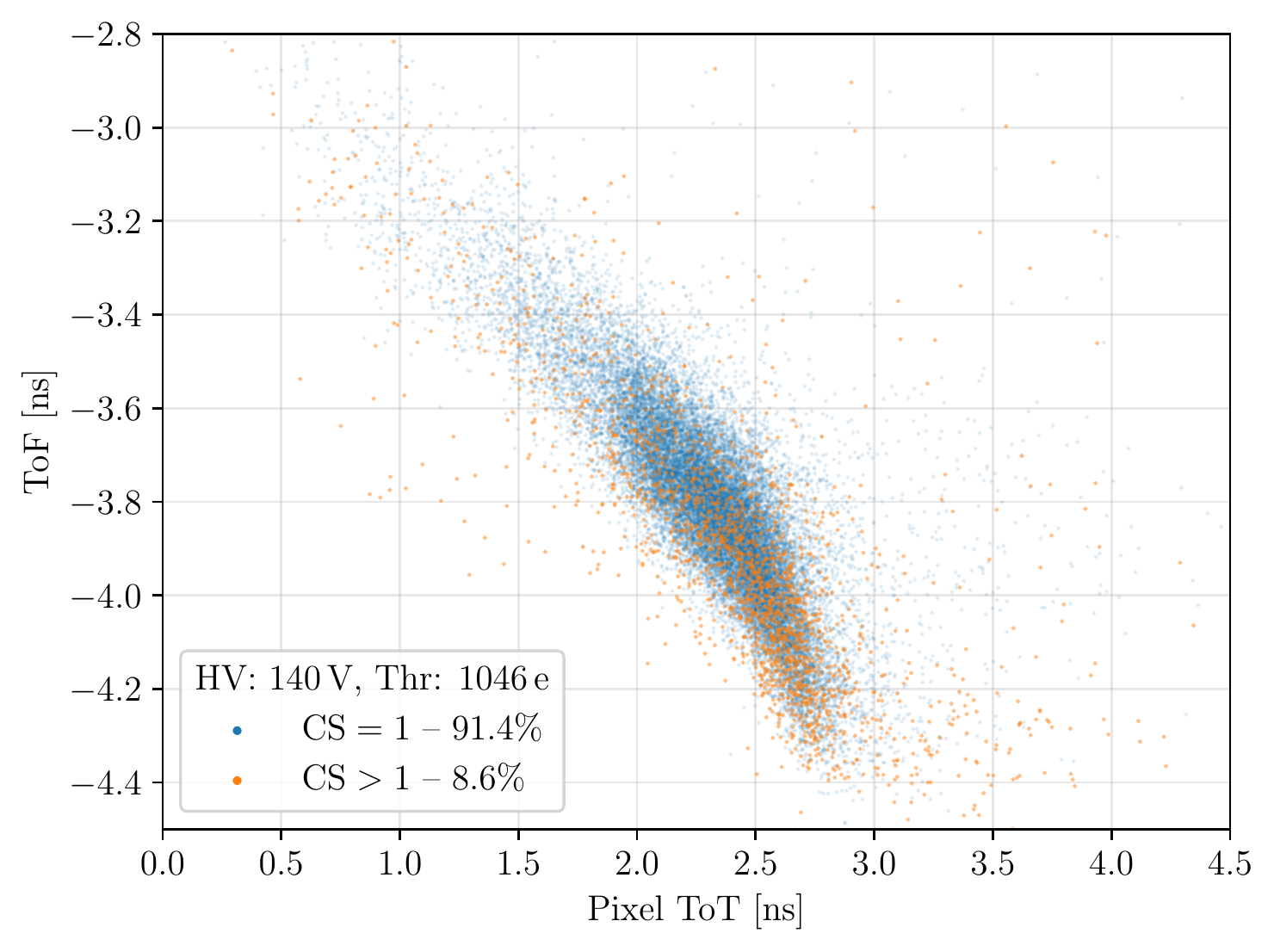}
	\caption{Difference of the pulse ToA at the oscilloscope between the pixel under test and the reference LGAD detector as a function of the ToT of the signal from the pixel under test. The data refer to a sensor bias voltage of 140 V, an amplifier current of \SI{50}{\micro\ampere} and a charge threshold of \num{1046} electrons. Blue dots represent events with cluster size one, orange dots events where at least one of the two adjacent pixels fired.}
	\label{fig:tw_correction}
\end{figure}

\begin{figure}[h]
	\centering
	\includegraphics[width=0.65\textwidth]{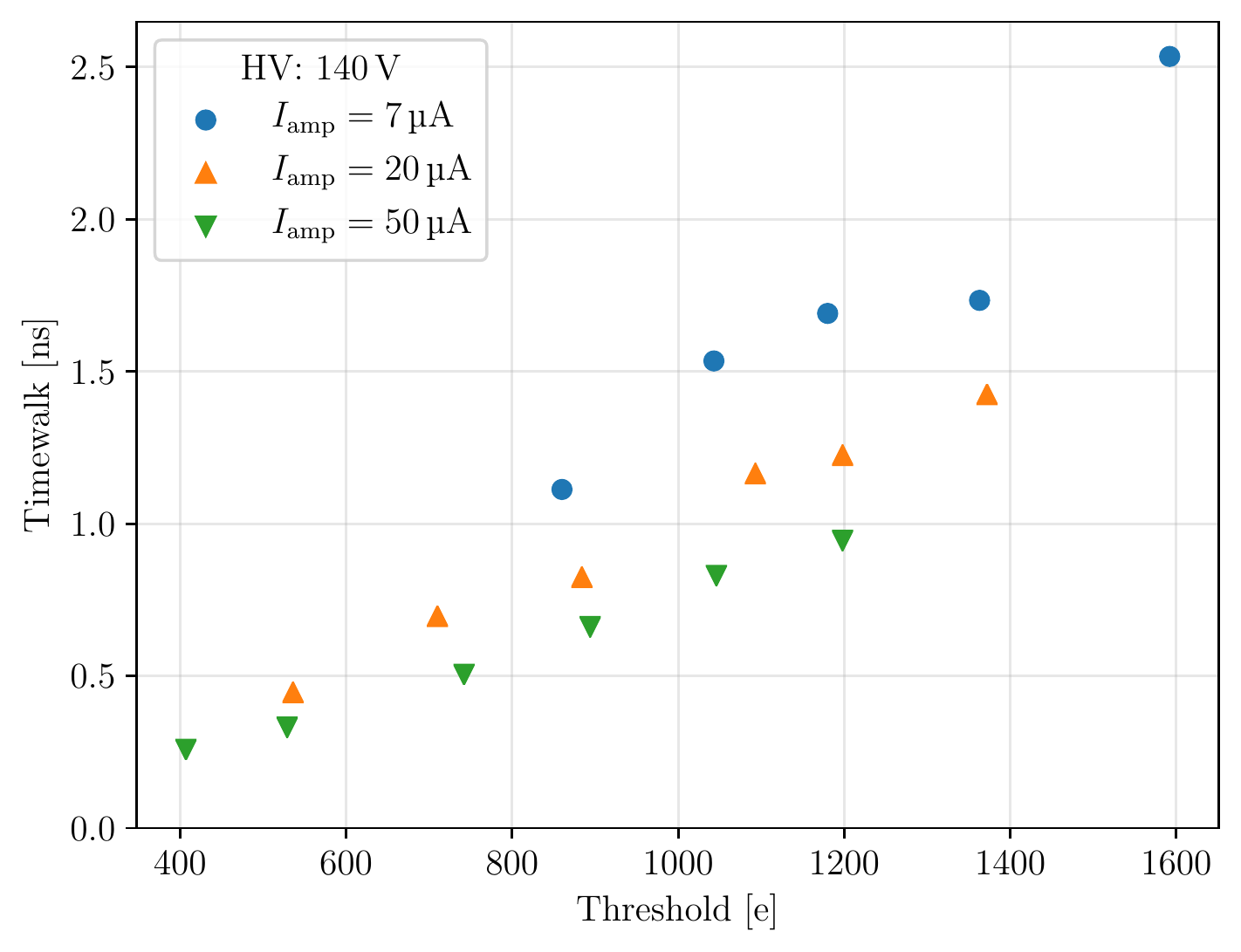}
	\caption{Maximum time-walk correction at different charge threshold values for the amplifier working points. For each threshold the range of time-walk correction was estimated as the width at 10\% of the amplitude of the distribution of $ \mathrm{ToF_{before~correction}-ToF_{after~correction}} $.}
	\label{fig:tw_correction_summary}
\end{figure}

For each threshold, the ToF was measured with the oscilloscope as the difference between the ToA values of the signals from the prototype under test and the reference LGAD. The time-walk can be partly corrected by using the distribution of the correlation between the ToF and the prototype-signal ToT.
As an example,
 Figure~\ref{fig:tw_correction} shows the TOF vs. TOT distribution obtained at a sensor bias voltage of 140V for an amplifier current of \SI{50}{\micro\ampere} and a threshold of 1046 electrons. The range of the time-walk correction depends on the discriminator threshold and on the amplifier current, as shown in Figure~\ref{fig:tw_correction_summary}, with values ranging from \SI{0.25} to \SI{2.5}{\nano\second}.

\begin{figure}[h]
	\centering
	\includegraphics[width=0.6\textwidth]{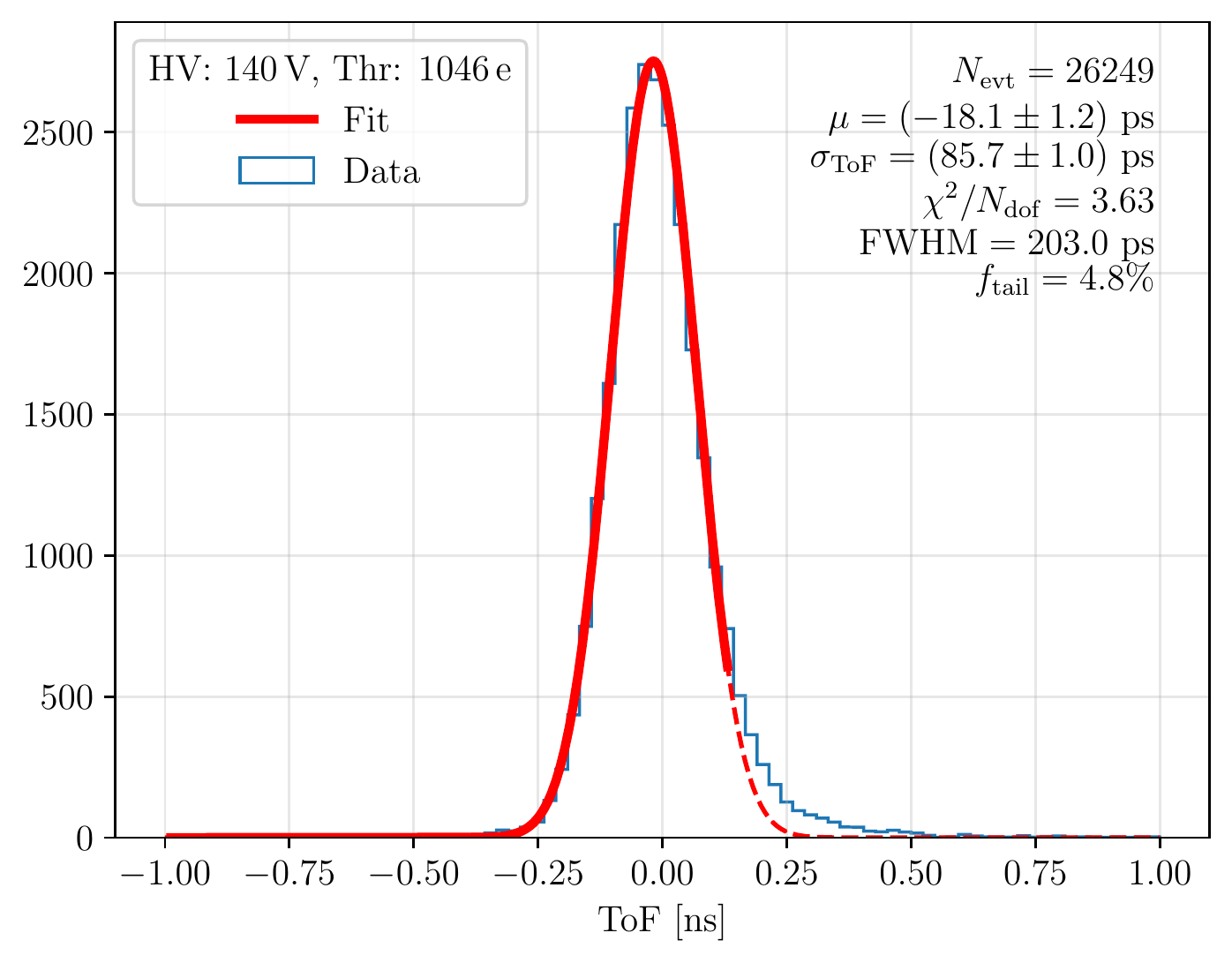}
	\caption{Distribution of the ToF between the pixel under test and the LGAD for an amplifier current of \SI{50}{\micro\ampere} and a discriminator threshold of \num{1046} electrons. The time jitter was estimated by a Gaussian fit - represented by the full red line - that excludes the tail of the distribution for ToF>\SI{125}{\pico\second}. The dashed line shows the continuation of the fitted Gaussian function outside the range of the fit. The fraction of events in the tail exceeding the Gaussian fit is $ f_{tail}=4.8\% $.
	The average value of the ToF was set close to zero by the time walk correction. }
	\label{fig:tof_hist}
\end{figure}

Figure~\ref{fig:tof_hist} shows the ToF distribution after time-walk correction for the same amplifier current and discriminator threshold of Figure~\ref{fig:tw_correction}. The mean value of the distribution was set close to zero by the time-walk correction. The time jitter of the distribution was estimated by a gaussian fit that excluded part of the tail at positive values of ToF that is caused by a systematic error in the time-walk correction~\cite{hexa_50ps}. 
Indeed, due to its non-linear transfer function, the discriminator produces an asymmetrical jitter of the ToT, sometimes overestimating its value. When this happens, the signal ToA values are under-compensated for time walk, producing the non-gaussian tail observed in the ToF distribution. 
In all cases, the full width at half maximum (FWHM) of the distribution was found to be compatible with the $ \sigma_{ToF} $ obtained from the Gaussian fit. The fraction of events in the tails exceeding the Gaussian distribution is typically below 5\% of the total number of events.

\begin{figure}[h]
	\centering
	\includegraphics[width=0.8\textwidth]{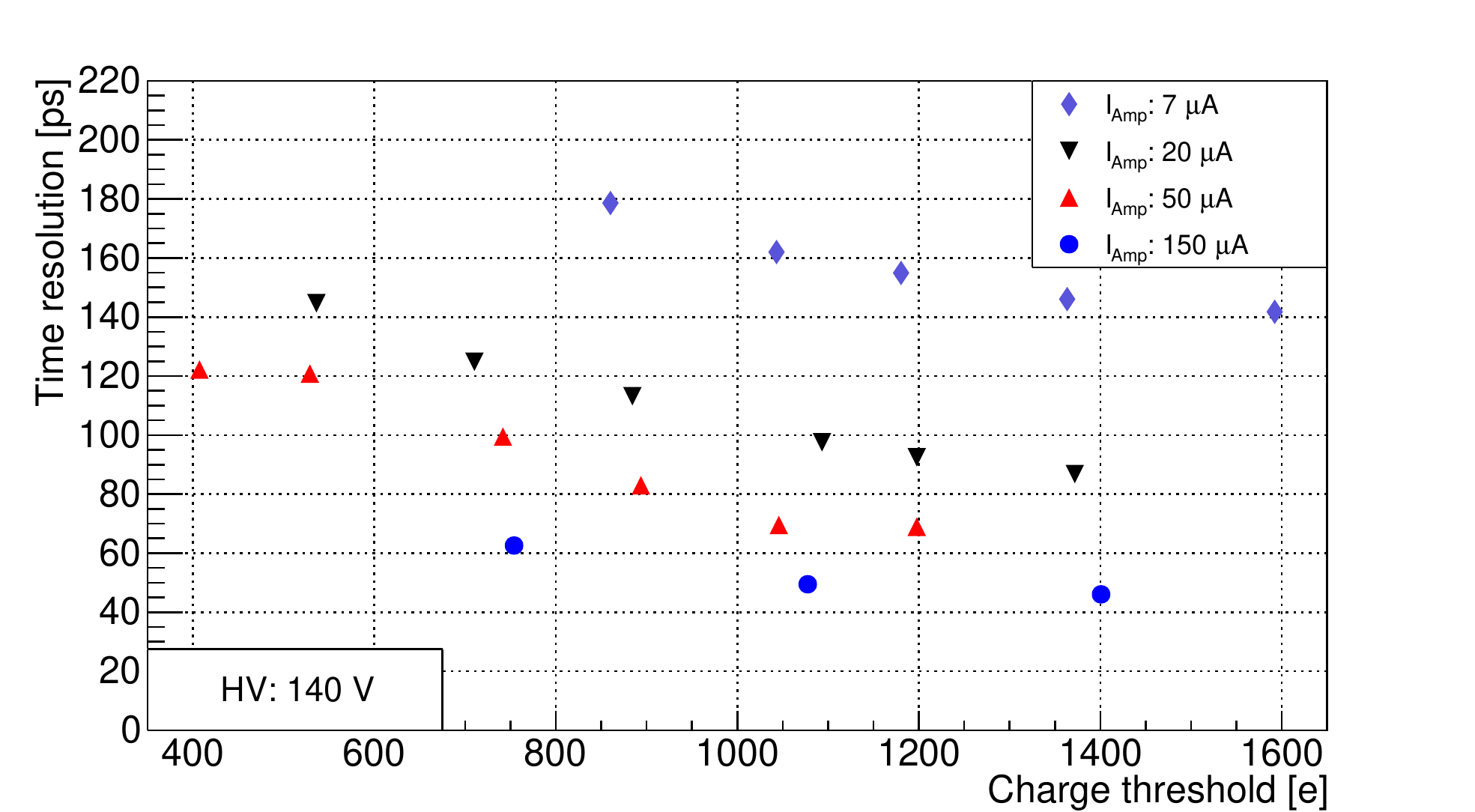}
	\caption{Sensor time resolution vs. discrimination threshold for amplifier currents of 7, 20, 50 and \SI{150}{\micro\ampere} and sensor bias voltage of \SI{140}{\volt}. The time resolution was obtained correcting for the time walk using the ToT at the output of the discriminator. The contribution of \SI{50}{\pico\second} from the reference LGAD to the time resolution was subtracted in quadrature from the ToF resolution. The results for an amplifier current of \SI{150}{\micro\ampere} are those from \cite{hexa_50ps}.}
	\label{fig:tof_hist_thr}
\end{figure}

The time resolution of the prototype under test with ToT-based time walk correction was finally estimated by subtracting in quadrature \SI{50}{\pico\second} (the time resolution of the LGAD) from the $ \sigma_\text{ToF} $ value obtained from the fit. Figure~\ref{fig:tof_hist_thr} shows the time resolution for the different amplifier currents as a function of the discriminator threshold. In all cases the measured time resolution is below \SI{180}{\pico\second} and improves at larger thresholds. Cadence Spectre simulations suggest that the dependence on the discriminator threshold indicates that the poor ToT resolution, which limits the accuracy of the time-walk correction, is the dominating factor for the sensor time resolution. The reduction of the amplifier current enhances this effect, increasing both the range of the time walk and the signal fall time, with consequent deterioration of the ToT resolution at the lower thresholds. Therefore, we conclude that the intrinsic jitter of the amplifier is small in this prototype with respect to the resolution of the time-walk correction for all amplifier currents considered.

\begin{figure}[h]
	\centering
	\includegraphics[width=0.8\textwidth]{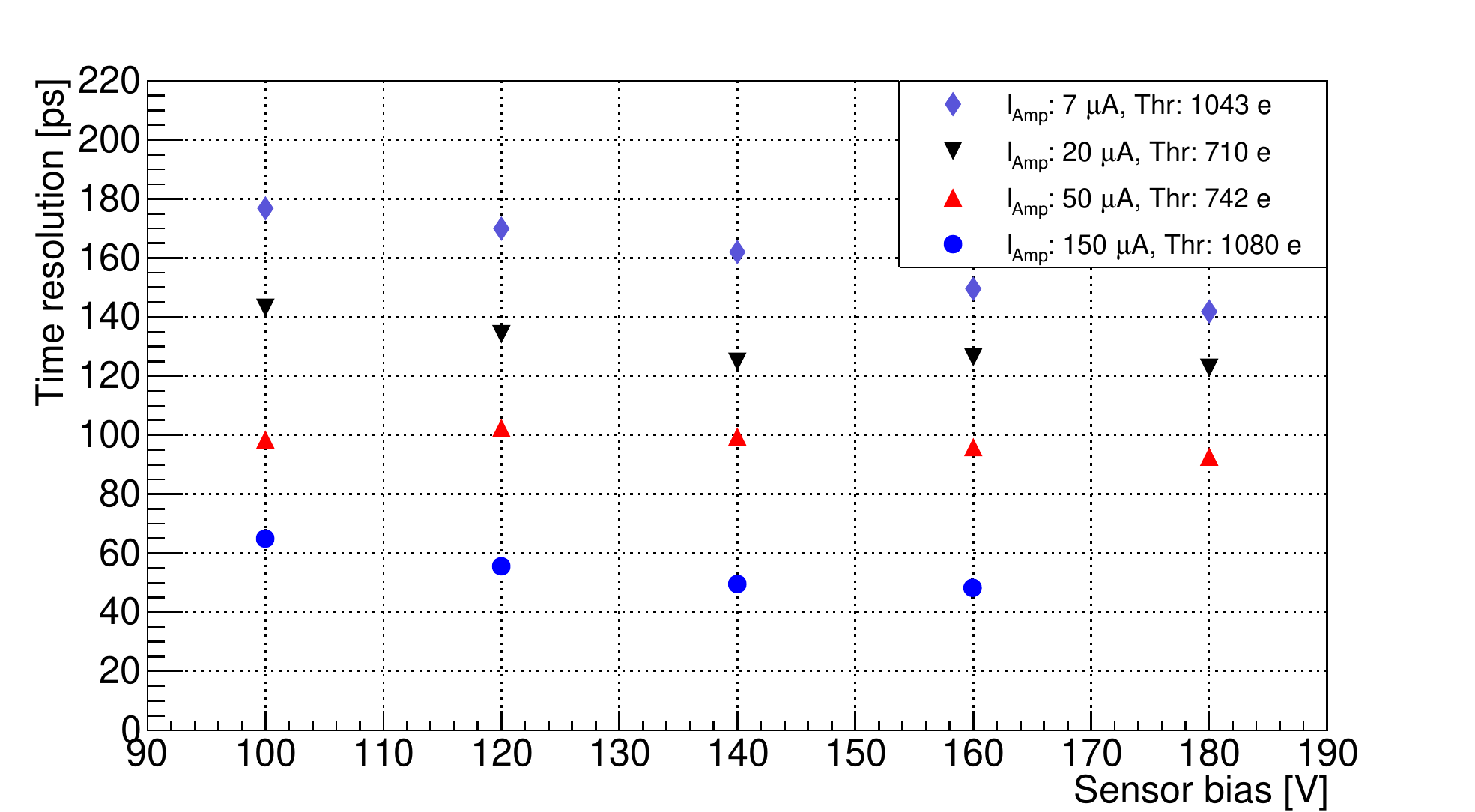}
	\caption{Sensor time resolution vs. sensor bias for amplifier currents of 7, 20, 50 and \SI{150}{\micro\ampere} and discrimination thresholds of 1043, 710, 742 and 1080 electrons, respectively. The results for an amplifier current of \SI{150}{\micro\ampere} are those from \cite{hexa_50ps}.}
	\label{fig:tof_hist_hv}
\end{figure}

Figure~\ref{fig:tof_hist_hv} shows that the time resolution improves at higher sensor bias voltages. This improvement can be explained by two factors: 
\begin{itemize}
	\item An increase of the depletion depth, which is associated to a smaller pixel capacitance and a larger charge signal from the MIPs (up to 13\% at the highest voltage). As a consequence it offers a better signal to noise ratio for the time walk correction.
	\item An increase of the average electric field in the active volume, with faster charge collection and improved intrinsic timing performance of the pixel sensor.
\end{itemize}
The contribution from the second effect is small and it does not depend on the amplifier working point, therefore the large improvement at the lowest current can be attributed solely to the increase of the signal charge.

\section{Conclusions and discussion on the impact of SiGe HBT in timing sensors}
\label{sec: conclusions}

The timing performance of a small-size monolithic silicon pixel prototype featuring an amplifier realised with SiGe HBT was measured  with a \Sr{} source setup.
For a pixel capacitance of 70 fF, a time resolution of \SI{140}{\pico\second} was achieved for an amplifier current of \SI{7}{\micro\ampere}, while \SI{45}{\pico\second} were measured at \SI{150}{\micro\ampere}. 
These excellent results show that the low noise and high transition frequency of SiGe HBTs can be used to produce amplifiers for silicon pixel detectors with low time jitter at very low power consumption. 
 
An analysis of the results indicates that the measured time resolutions are limited by some design characteristics of this prototype and not by the performance of the SiGe HBT. In particular:
\begin{enumerate}
	\item A study of the sensor response with a CADENCE Spectre simulation suggests that the improvement of the timing performance for higher thresholds is indicative of a time resolution limited by the use of the measured ToT to compensate for the signal time walk. 
	\item The improvement in timing performance at higher sensor bias indicates that an increase of the width of the depletion layer is beneficial, especially  at very low power consumption. 
	\item For the bias voltage and resistivity used here, the   intrinsic sensor contribution to time resolution from the charge deposition profile and the sensor uniformity of response is expected to be below \SI{30}{\pico\second}~\cite{werner}.
\end{enumerate}



\begin{figure}[htbp]
	\centering
	\includegraphics[width=0.7\textwidth]{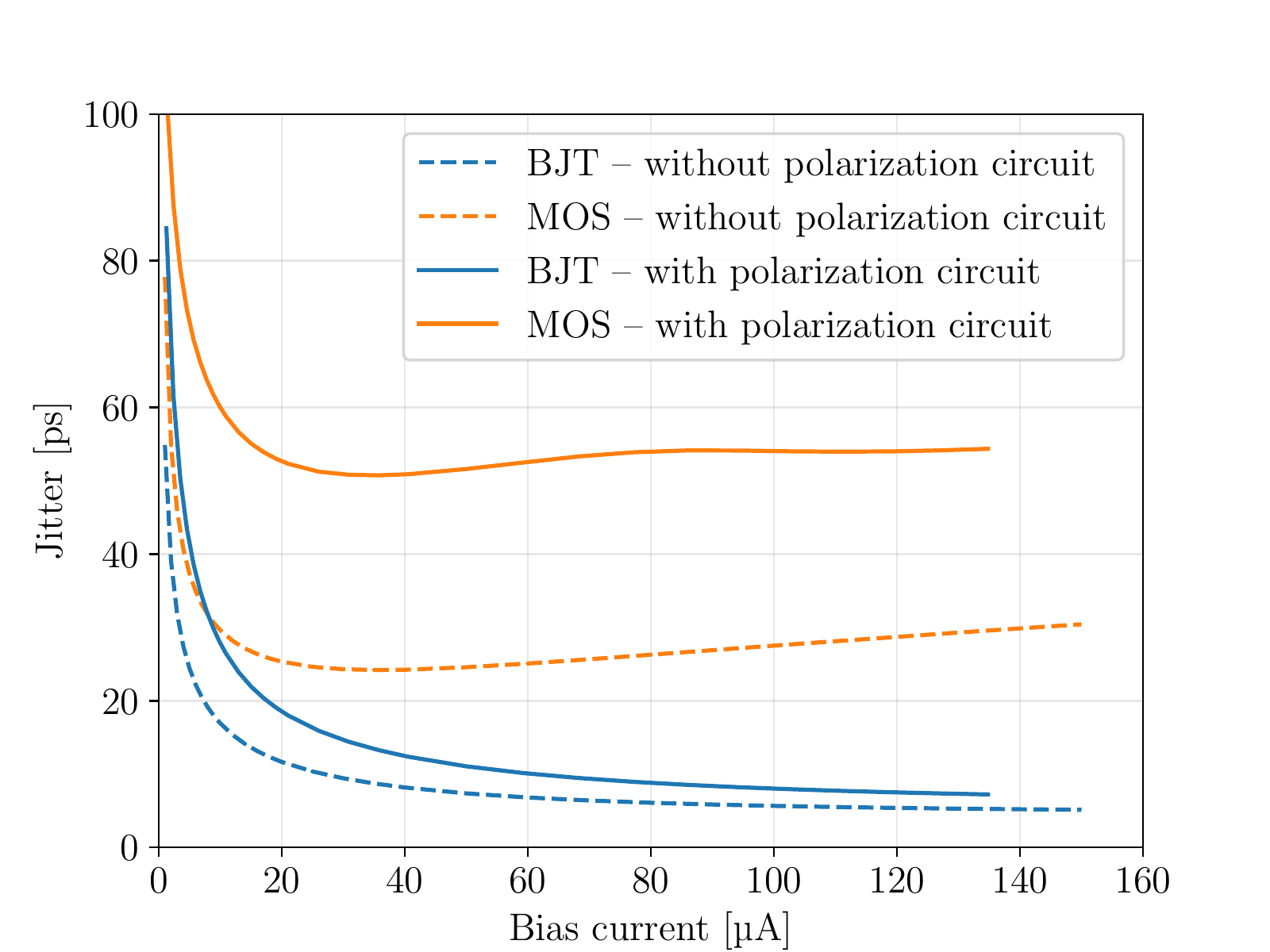}
	\caption{CADENCE Spectre simulation of the time jitter as a function of the supply current for an amplifier based on a CMOS transistor in common source configuration (orange) and for a SiGe HBT in common emitter configuration (blue). The simulation was performed for an  input capacitance of \SI{80}{\femto\farad} and an input charge of \num{1600} electrons. The time jitter was defined as the ratio between the voltage noise at the output of the amplifier and the maximum slope of the  rising edge of the pulse. Dashed lines: biasing circuit made of ideal, noiseless resistors. Full lines: realistic CMOS-based biasing circuit.}
	\label{fig:simLOAD}
\end{figure}

To quantify the ultimate performance of this technology, a single-transistor amplifier in common source (emitter) configuration was simulated using CADENCE Spectre.
The simulation investigates the trade-off between the intrinsic timing jitter and the power consumption for a minimum-size NMOS transistor and a SiGe HBT in the same SG13G2 130nm BiCMOS process of IHP Microelectronics. 
Figure \ref{fig:simLOAD} shows the results for two scenarios:
a noiseless, ideal polarization of the transistors with a parasitic capacitive load of \SI{2.5}{\femto\farad}, and a realistic CMOS-based polarization circuit with the same capacitive load. 
In both cases a pixel capacitance of \SI{80}{\femto\farad} was considered.
The simulation shows that the HBT provides a lower intrinsic jitter than the CMOS transistor by more than a factor of two, despite the extra capacitance that was added to the standard HBT in the insulation process. This superior performance is further enhanced when the robustness to parasitic capacitance of the HBT is taken into account.

\acknowledgments

The authors wish to thank the technical staff of the Department of Particle Physics of the University of Geneva for the assembly of the instrumentation and for the support in preparing the test, and our colleague Nicol\`o Cartiglia for providing the FBK reference LGAD.

\newpage

\bibliographystyle{unsrt}
\bibliography{bibliography.bib}

\end{document}